\documentclass[12pt]{article}

\usepackage{amsmath}
\usepackage{hyperref}
\usepackage{authblk}

\usepackage{todonotes}

\begin{document}

\title{Shape-based event pileup separation in Troitsk nu-mass experiment}
\author[1]{Vasily Chernov}
\author[1,2]{Alexander Nozik}
\affil[1]{Institute for Nuclear Research RAS, 117312 Moscow, Russia}
\affil[2]{Moscow Institute of Physics and Technology, Dolgoprudny, Moscow region 141700, Russia}

\maketitle

\begin{abstract}
    Nowadays continuous signal digitization becomes a standard  procedure in experimental physics. Though, signal pileup separation at high count rate remains a problem. The article presents algorithms for detecting and extracting events  based on  shape of a single pulse. An example of applied use of the described algorithms is also given on the example of the Troitsk nu-mass experiment.
\end{abstract}

\section{Introduction}
One of the major problem at the Troitsk nu-mass experiment in search for sterile neutrino with mass up to 5 keV (\cite{2015JInst..1010005A}) is the high count rate on the main detector. According to proposal, it should handle up to 40-50 kHz count rate with good discrimination of pile-up and very precise dead time estimation (down to 100 ns precision or even lower). With signal shape width of about 4 $\mu s$, it is hard to implement hardware signal separation. So it was decided to use continuous digitization and perform the signal separation in off-line mode. The current setup includes very cheap RudShel Lan10-12PCI ADC board \cite{Lan10-12PCI}, which have maximum precision of 20 ns. Due to specifics of board oscilloscope mode it appeared that the single event mode could not be used on high count rates. Instead, we acquired continuous fixed length blocks, performed fast zero suppression in on-line mode and left everything else for off-line analysis. This acquisition mode allows to partially bypass board hardware limitations and eliminates hardware dead time, but requires a software to perform off-line pulse separation.

There are several ways to detect event pileup. For example, restoring the shape of a spectrum using a neural network. The algorithm described in \cite{2017JHEP...12..051K} uses modeled spectra as a training sample and requires knowledge of the form of the original, undistorted spectrum. Another way is described in \cite{2013NIMPA.717...21C}. It is based on analytical model that predicts dead time and spectrum distortions caused by signal pileups. Despite good results, the model is rigidly tied to the analytical pulse formula and therefore is not applicable in other systems without significant modification.

Our proposed correction algorithm is also based single event shape.
However, instead of using information received from outside, such as shape of the spectrum or the analytical signal formula, it relies only on the continuous signal data collected from the ADC, and extracts all the necessary parameters from the data itself. This makes the algorithm applicable for various signal digitization problems without significant changes.


\section{Hardware}
Lan10-12PCI has a memory buffer of 512 kBit and can hold only one frame. The restriction in the simultaneous storage of only one frame in the internal memory makes it necessary to dump board memory to PC after each frame. With Lan10-12PCI constant dump time of $55 ms$ it leads to impermissible dead time for trigger-based acquisition mode. However, board allows to acquire frames by the program trigger which, with the maximum allowable frame size of 1048576 bins and the bin size equal $320 ns$, allows to receive continuous signal data with constant 87\% live time. Considering the signal rising time of 2 $\mu $, the mode is suitable for Troitsk nu-mass setup. Better hardware could produce better sampling rate and live time.

\subsection{Zero-suppression}
\label{zero-suppression}
To save hard disk space, continuous signal data frames are preprocess by the zero-suppression algorithm. The algorithm truncates bins which have an amplitude below the threshold and don't have bins above the threshold in a some neighborhood. Thus, a continuous signal frame is divided into variable-length subframes, with one or more events contained each.

The size of the subframes should be chosen based on the event after-pulse decay rate - it must be greater than length of the maximal possible after-pulse in dataset. With this restriction, algorithm will safely separate frame
to independent subframes, since the impact of the closest event to the right of the subframe will be obviously less than the minimum size of truncated area.

\begin{figure}
	\includegraphics[width=0.9\textwidth]{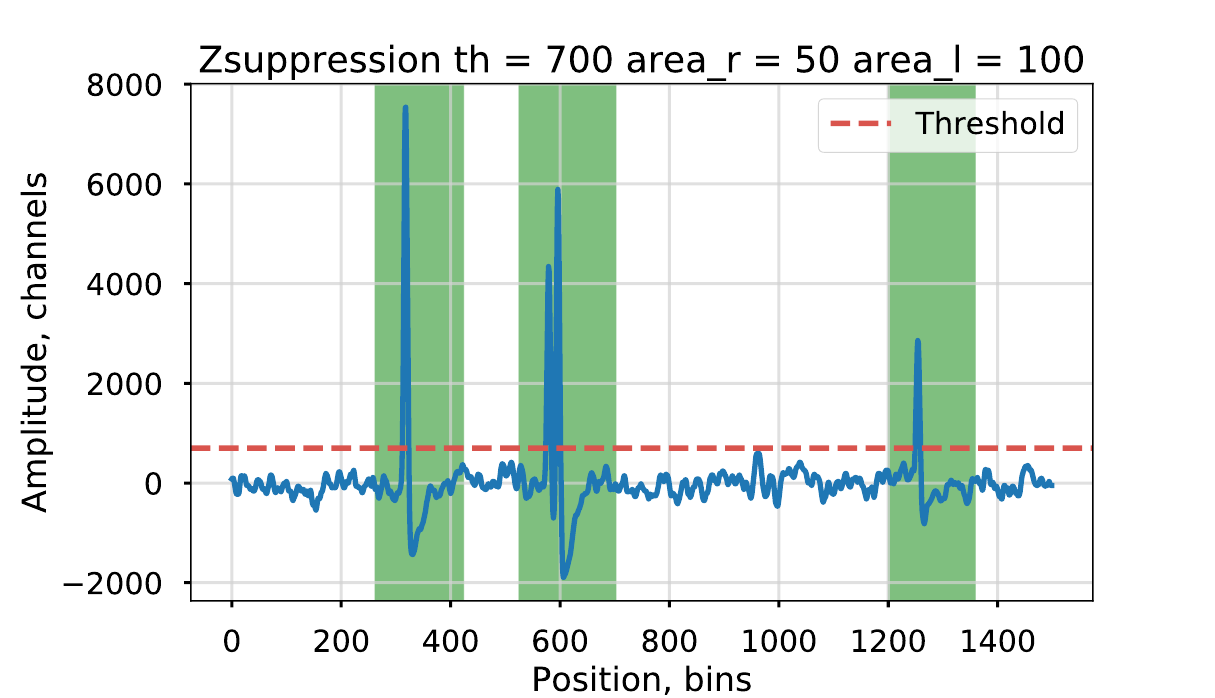}
	\caption{Zero-suppression algorithm work example. Highlighted areas correspond to algorithm output subframes. In this case they contain 1, 2 and 1 events respectively. The rest of the continuous signal frame is truncated. }
\end{figure}

The zero-suppression operation algorithm:
\begin{enumerate}
	\item Create frame binary mask
    \begin{enumerate}
    	\item Creation of threshold mask 
    	\item The application of morphological dilation with a core corresponding to the desired boundaries of the event
    \end{enumerate}
	\item Extract subframes passed by mask
\end{enumerate}

This implementation avoids duplication of events boundary areas due to merging of close events into one subframe. Further in the work, for simplicity, we will refer to the subframe, resulting from the zero-suppression preprocess as a simple "frame" and "continuous signal frame" - for Lan10-12PCI input frame before preprocess.

\section{Continuous signal simulation}
To develop and test extracting algorithms, a large labeled continuous signal data sample is required. The only way to obtain it is simulation. During the work, we developed Lan10-12PCI board data generator, that produces continuous signal frames simulating real device output. To create each continuous signal frame it uses set of event defining parameters such as it amplitude and time. Since the parameters are defined explicitly the output becomes labeled automatically.

Generator output is developed to be fully compatible with real board raw output. This choice has its advantages and disadvantages. On the one hand raw board output is the primary input for the entire processing system, the use of such generated marked data allows to test all processing steps. In addition, there is no need to develop excess testing only code, because all used data manipulations will be required for working with real board. Another advantage is the ability of developing without the board itself, since the generator can also be used as a virtual device. The disadvantage is the need to perform all dependent steps before testing a specific part of the processing. This greatly reduces testing speed, especially for the final processing steps, compared to the case of generating data for specific step. Also, a bottleneck could appears in case when some intermediate steps works much longer than the tested one.
Weighing all the pros and cons the method of generating raw board input was chosen as the optimal in terms of development costs. \\
Architecturally, the data generator consists of two modules:

\begin{itemize}
    \item Noise generator;
	\item Clean event shape generator.
\end{itemize}

Continuous signal is obtained by overlaying clean event shapes with noise background.

\subsection{Noise simulation}
We assume that noise consists of a large number of low amplitude events uniformly distributed in time and having the same shape as the shape of useful event. With this condition, the noise can be generated by explicit superposition of the shapes created by clean event shape generator without additional modules. However, to speed up simulation, it is possible to optimize noise generation.

Taking into account the high frequency and peaks distribution of noise events, it is possible to generate superimposed noise events final shape with method based on applying a composition of smoothing operations on a random numbers array.\\
Noise generation algorithm then works as follows:
\begin{enumerate}
	\item Create random numbers array with uniform distribution. Array size is equal to output frame size.

	\item Convolve array with a large constant core to achieve average number of peaks similar to real noise. Coarse tune amplitude distribution.

	\item Convolve array twice with small constant cores to achieve necessary blurring of small peaks. Fine tune amplitude distribution.

	\item Round array values to integer and shift them by 2 bits to make them  be similar to 12 bit bin values acquired by board.
	
\end{enumerate}

The method allows to generate superimposed shape of large number of uniformly and densely distributed over time events at low costs. Unlike an events superposition method, result won't be labeled but it is acceptable for noise.

Comparison of parameters of real and simulated amplitude spectra was used for noise generator validation. To obtain a real noise spectrum the frame slices from left border to frame first event beginning was used. Due to zero-suppression applied to data and while fulfilling the constraint from the procedure explained in section \ref{zero-suppression} they will be the only source of undistorted real noise data. 

Convolution kernel sizes were tuned manually. During the tuning we watched coincidence between real and generated noise for:

\begin{itemize}
	\item Correspondence of amplitudes spectrum of real and generated noise.
	\item Visual similarity of the shapes of noise peaks and average distances between neighboring peaks.
\end{itemize}


\begin{figure}
	\includegraphics[width=0.9\textwidth]{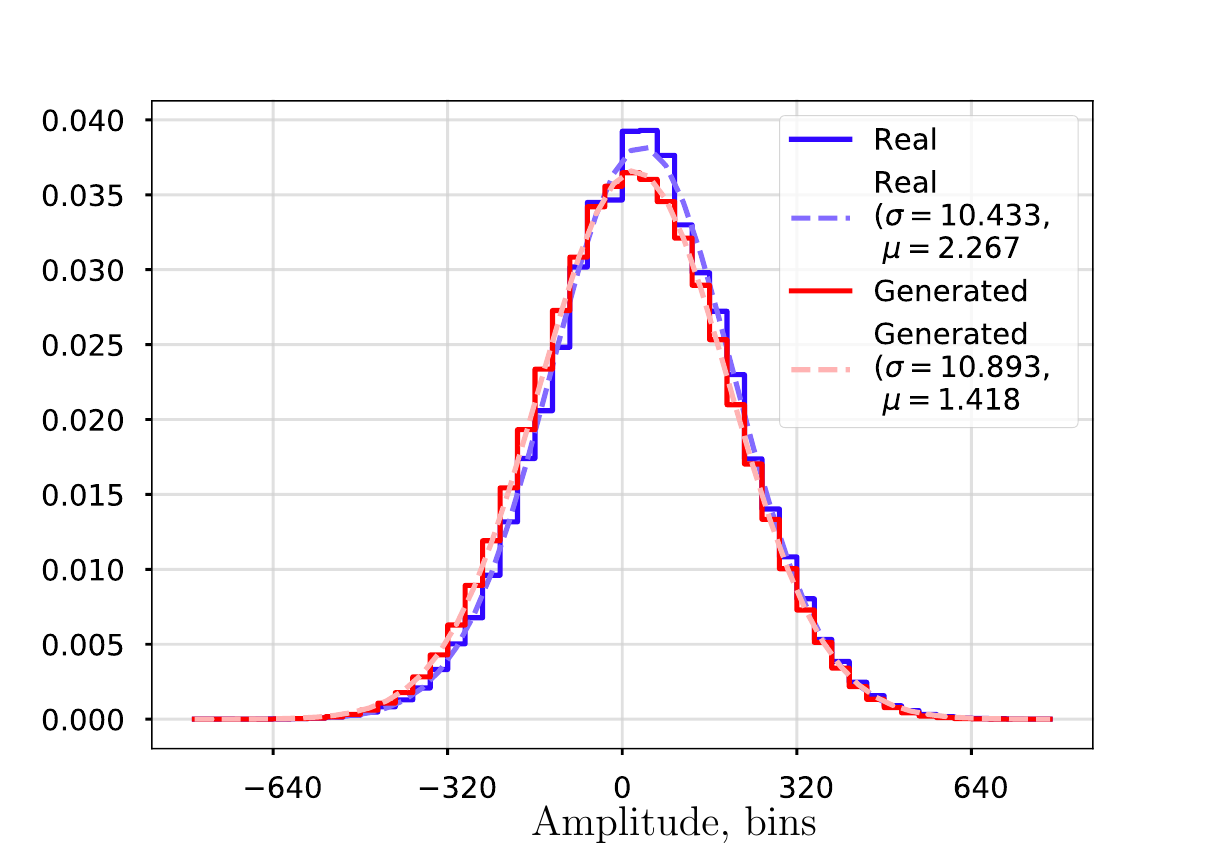}
	\caption{Noise spectrum comparison simulated vs real.}
	\label{noise-generator-validation}
\end{figure}

The spectra peaks at Fig.~\ref{noise-generator-validation} show that the baseline for Troitsk setup fluctuates slightly between series of measurements. Testing was conducted on later sets of data and mean for noise has shifted by 12 channels. However, the statistical parameters of the noise are close to each other.

Now we know the mean and the standard deviation for the distribution of the noise amplitudes, which we need when analyzing the form of events.

\subsection{Event shape modeling}
To model an event, one need to know its clean form without noise. Forms of events can be extracted from the data by grouping events by amplitude value and then averaging the shapes by group. The frame size after trimming also gives a preliminary condition which can filter out frames with overlapped events.

\begin{figure}
	\includegraphics[width=0.9\textwidth]{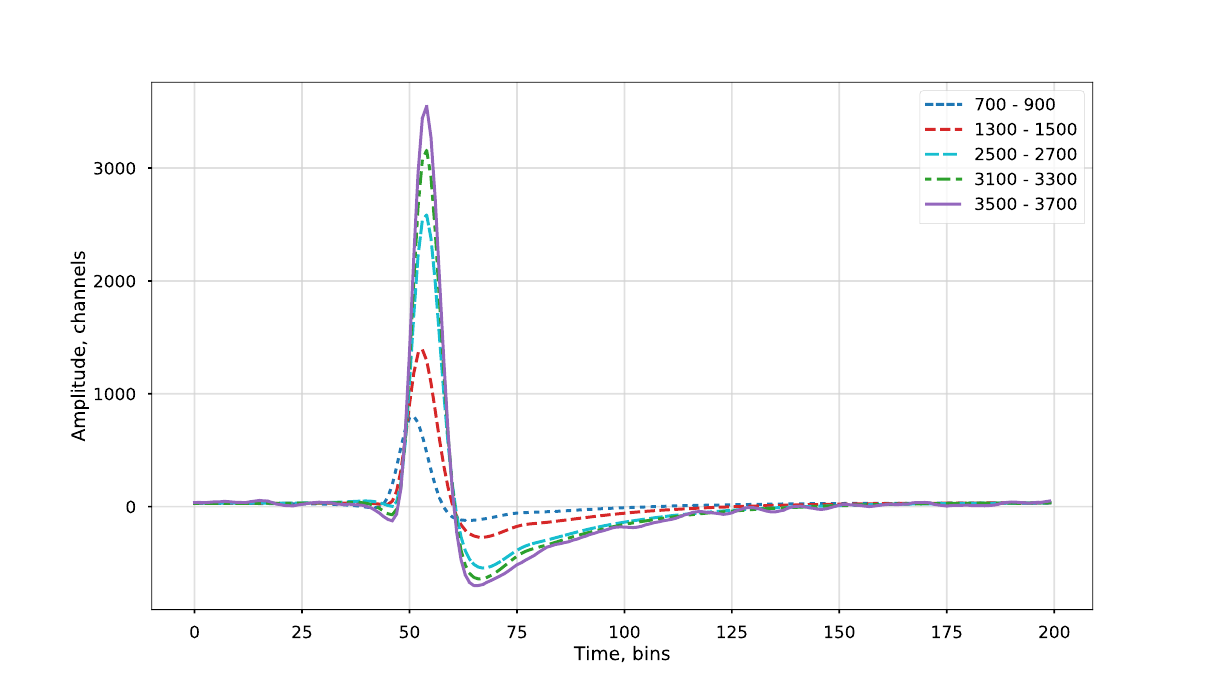}
	\caption{Averaged event shapes for groups of amplitudes. Horizontal axis bin size is 320 ns.}
	\label{averaged-shapes}
\end{figure}

For the Troitsk nu-mass dataset averaged shapes shown on figure \ref{averaged-shapes}. The following shape properties can be concluded from the figure:

\begin{enumerate}
	\item The peak of the event can be approximated by a Gaussian.

	\item Event after-pulse decays exponentially.

	\item The ratio between peak and after-pulse amplitudes is preserved.

	\item The distance between the positions of the peak maximum and the after-pulse minimum is preserved.
\end{enumerate}

To describe the obtained forms of real signals, an analytical formula has been found:

\begin{equation}\label{shape_formula}
\begin{split}
    t_f = 2.9604, p = 2.2056, t_a = 0.3701, \sigma = 0.3416, f_b = 3.125e+6, \\
    g(x) = exp(\frac{|\sigma f_b log(x)|^p}{2}), g_r(y) = \frac{-2 log(y)^{1/p}}{\sigma f_b}, \\
    s(y) = 
    \begin{cases}
      (((1 + 2 x f_b s)^t_f - 1)  exp(-x f_b s))^{-1}, x > 0 \\
      0, x \leq 0
    \end{cases}, \\
    signal(x, amp, pos) = g(x - pos) + s(x - g_r(0.1) - pos) t_a amp.
\end{split}
\end{equation}
where: \\
$g(x)$ - Gaussian function, $g_r(y)$ - reverse Gaussian function, $s(x)$ - after-pulse function,  $signal(x, amp, pos)$ - signal function.

The signal front is completely coincides with the Gaussian function. After-pulse function starts at point where Gaussian amplitude reach value equal to 10\% of the maximum. Parameters were chosen by fitting.

The analytic signal formula is used to increase the performance, however, separation algorithm doesn't require it - event shapes given numerically are sufficient.

\begin{figure}
    \includegraphics[width=0.45\linewidth]{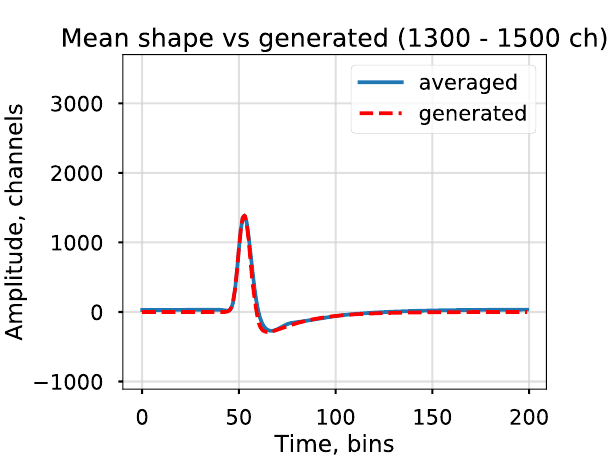} 
    \includegraphics[width=0.45\linewidth]{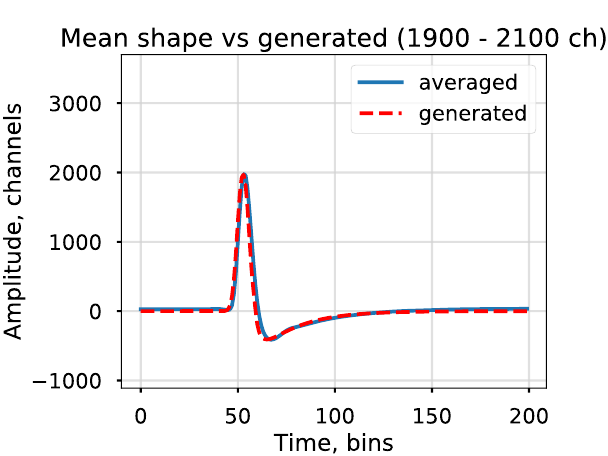} \\
    \includegraphics[width=0.45\linewidth]{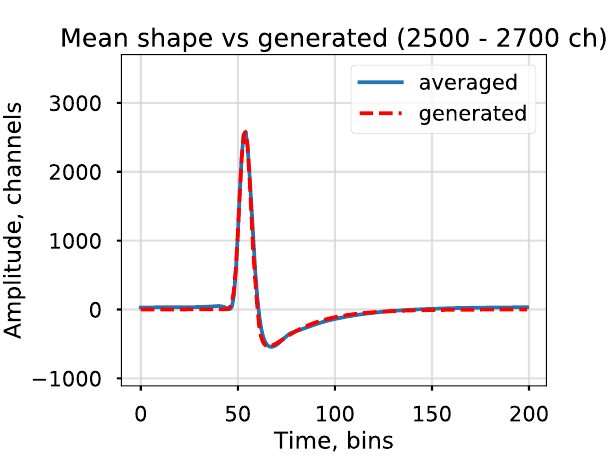} 
    \includegraphics[width=0.45\linewidth]{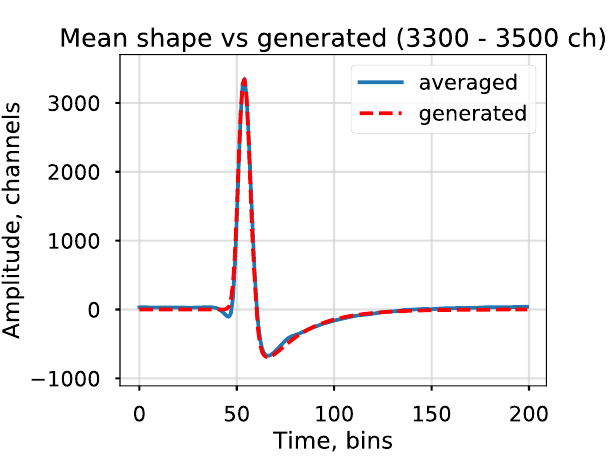} \\
\caption{Real averaged vs generated signal shapes.}
\label{events-shape-comparison}
\end{figure}

Fig.~\ref{events-shape-comparison} show comparison of generated shapes with shapes obtained by averaging real events. In this comparison, the real averaged pulse maximum and its position was used as amplitude and time of corresponding generated event.

\section{Events extraction algorithms}
Several selection algorithms have been investigated and tested. Each algorithm, has been tested on generated file with duration of 300 seconds and a count rate of 40 kHz.

To define the correspondence between the original events and the events obtained as a result of the algorithm operation, the following rules were used:

\begin{itemize}
	\item Event corresponds to the original event if:
    \begin{itemize}
    	\item original event is the closest in time,
    	\item the distance between the original and the extracted event is in range of 3200 ns (10 time bins).
    \end{itemize}
	\item Original event is considered to be unrecognized there is no extracted event within the range of 3200 ns.

	\item Extracted event is considered to be false positive if there is no one original event within the range of 3200 ns.

	\item Extracted event is considered to be pileup if two or more original events correspond to it.
\end{itemize}

The following metrics were used for testing:

\begin{itemize}
	\item Algorithm running time;

	\item Number of false positives;

	\item Number of unrecognized original events
\end{itemize}

The efficiency of the algorithms is compared by effective dead time.

\subsection{Method 1. Simple}
The first method is the simplest, and consists in isolating local extrema above the threshold. To separate close events, the algorithm requires a clear bend between the peaks. This method is the fastest, however, due to its simplicity, it require events to be far from each other to successfully split them. Also, the algorithm does not correct the amplitudes of subsequent events by the tail value of previous event.

\begin{figure}
\centering
    \includegraphics[width=0.45\linewidth]{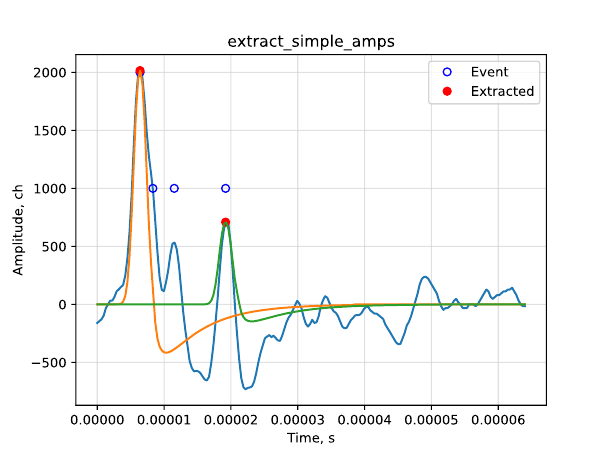} 	\includegraphics[width=0.45\linewidth]{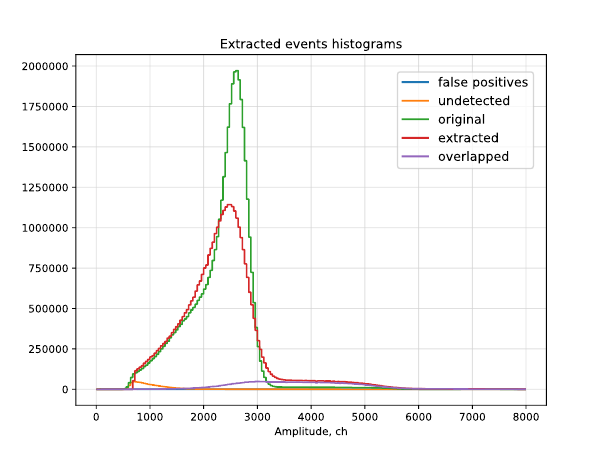} \\
    \includegraphics[width=0.45\linewidth]{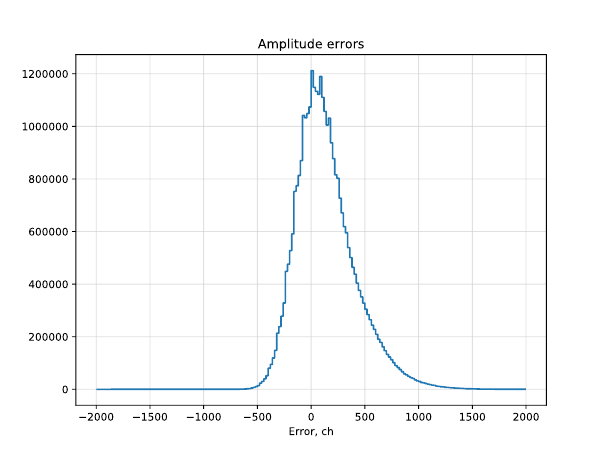} 	\includegraphics[width=0.45\linewidth]{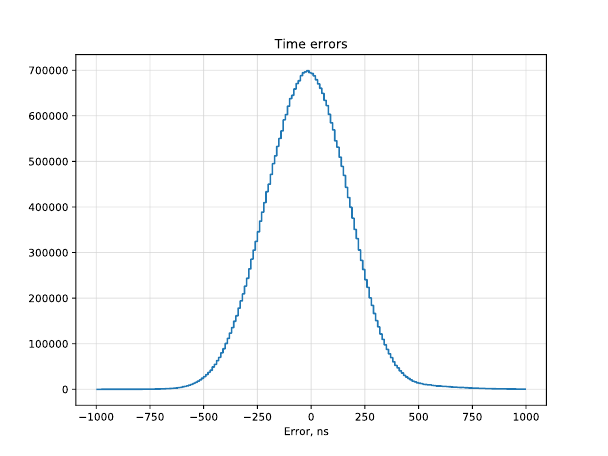} \\
    \caption{Method 1 testing results. The Upper left graph shows the extraction results on single example frame. Top right graph contains amplitude histograms for original, reconstructed, unrecognized, and pileup events. Bottom left and bottom right graphs shows amplitude and time error distributions for reconstructed events respectively.}
    \label{method-1-metrics}
\end{figure}

The graphical results of tests are shown at Fig.~\ref{method-1-metrics}. The algorithm did not recognize the second and third events of the test sample: the second - because of the lack of inflection between the peaks, the third - due to the imposition of the event on the minimum of the tail of previous event. The position of the fourth event has been successfully recognized, but it's amplitude has an error caused by overlapping previous events after-pulses. The contribution to errors caused by the lack of amplitude correction can also be seen on the amplitude errors histogram.

\subsection{Method 2. Approximate After-pulses}
The second method allows, with a relatively small increase in complexity (about 30\% ), to get rid of some of the disadvantages of the first method. It works as follows:

\begin{enumerate}
	\item Locate frame local maxima above the threshold.

	\item Process found peaks as a sequence:

    \begin{enumerate}
    	\item Preserve amplitude and position of the current peak;
    
    	\item Generate shape by amplitude and position;
    	
    	\item Subtract generated shape from frame (meaning all subsequent events);
 
    	\item Go to the next peak;
    \end{enumerate}
\end{enumerate}

Also, as in the first method, the detection of events depends on the bend of the signal above the threshold. For optimization purposes, peaks locating operation runs only once at the beginning of the frame processing. However, this method corrects the amplitudes of events that have arrived at the tail of previous events.

\begin{figure}[h]
    \centering
    \includegraphics[width=0.45\linewidth]{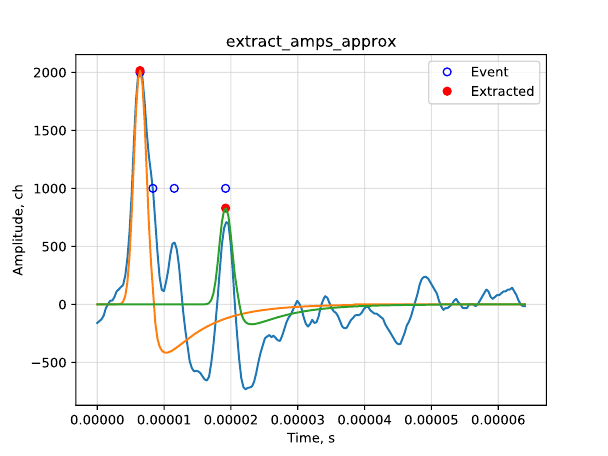} 	\includegraphics[width=0.45\linewidth]{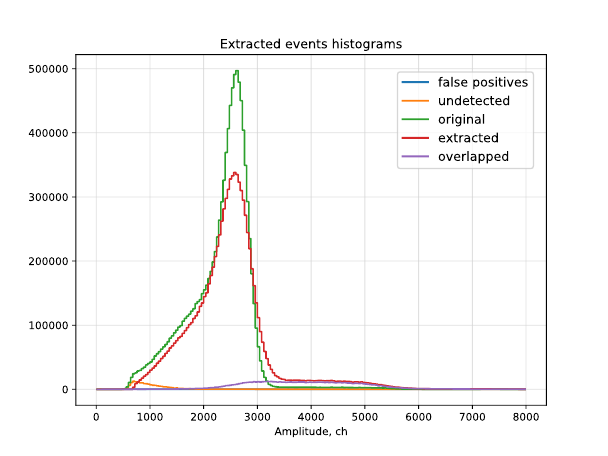} \\
    \includegraphics[width=0.45\linewidth]{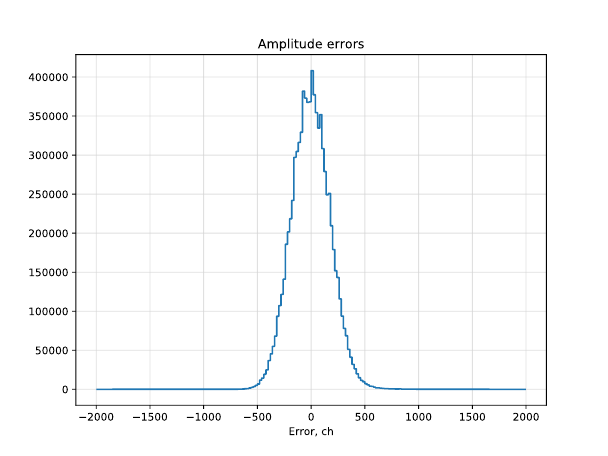} 	\includegraphics[width=0.45\linewidth]{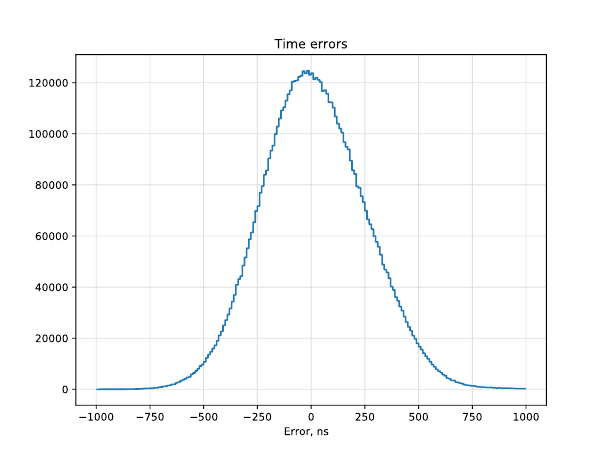} \\
    \caption{Method 2 testing results.}
    \label{method-2-metrics}
\end{figure}

Fig.~\ref{method-2-metrics} shows that the algorithm also failed to recognize the second and third events due to the overlapping with after-pulses. The fourth event amplitude is now adjusted by earlier events after-pulse, but also has an error caused by overlapping the tails of the second and third events which are not recognized. The amplitude errors histogram shows that the method adjusts amplitudes of reconstructed events affected by after-pulses.

\subsection{Method 3. Front fit}
The third method has two major differences. First one concerns peaks detection - in this algorithm the peak search is repeated every time the event shape is subtracted from the frame. After extracting the next event algorithm takes the left-most one from re-searched peaks to process.  
Second difference is change of peak position estimation - the position is fitted from several points in the front of the event. The fitting range is chosen in resect to events arrival time: the front of the event is not distorted by the subsequent events and since each time we take left-most event - it will be adjusted by after-pulses of earlier processed events from the frame. This approach allows to:

\begin{itemize}
	\item Resolve events that do not have a separation in the form of a bending.
	\item Extract events that initially have an amplitude below the threshold
\end{itemize}

\begin{figure}[h]
\centering
	\includegraphics[width=0.45\linewidth]{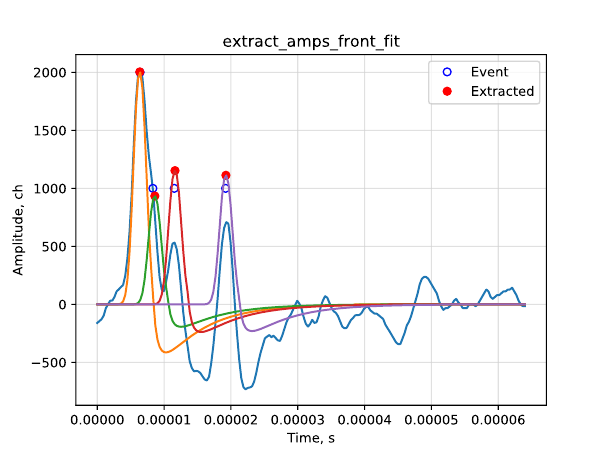}
    \includegraphics[width=0.45\linewidth]{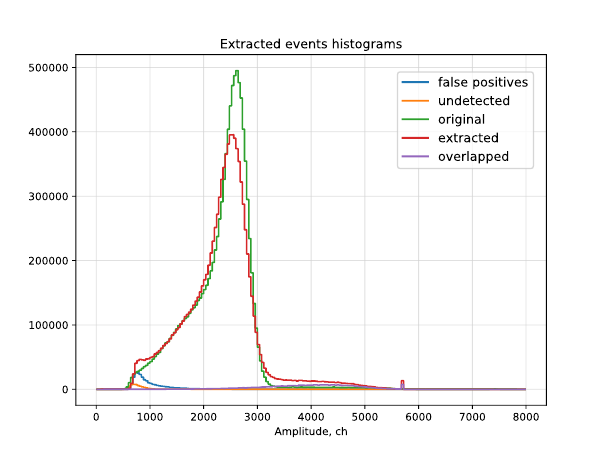} \\
    \includegraphics[width=0.45\linewidth]{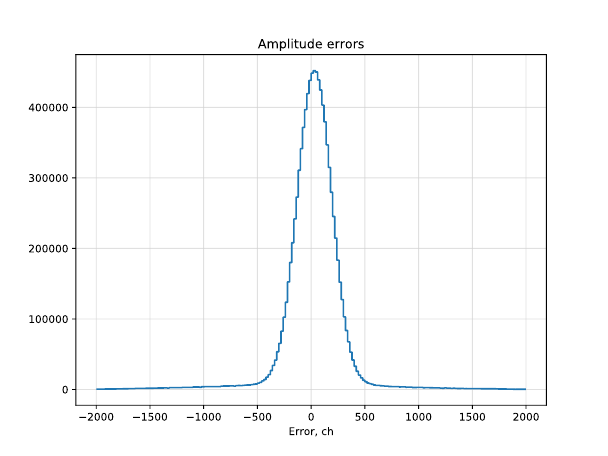} 
    \includegraphics[width=0.45\linewidth]{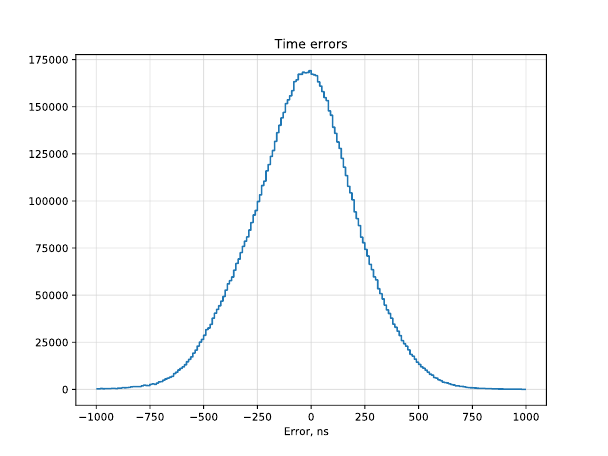} \\
\caption{Method 3 testing results.}
\label{method-3-metrics}
\end{figure}

Testing results are shown at Fig.~\ref{method-3-metrics}. The algorithm successfully resolves and adjusts amplitudes for all events. Errors in extracted peaks could be caused by noise and cumulative errors in recognition of previous events.

Also, it could be seen that peaks time extraction is less accurate than in first and second methods. This can be caused by a greater sensitivity to noise due to fitting by the front.

This method generates much greater number of false positives (2\%  vs. 0.005\%  in first and second). However the most of false positive events are located near the lower border of the amplitude spectrum and have amplitude values close to noise. This makes it possible to almost completely filter them out without large data loss by increasing threshold.

\subsection{Comparison of the methods}
The table \ref{comparison-summary} shows testing results of all three algorithms on the same generated file. The acquisition time is 300 s, the counting rate is 40 kHz. In this table, the ratios of false and pileup events are counted from the total number of actual recognized events. The effective dead time is defined as a dead time that causes the same amount of losses.

\begin{table}[h]
\centering
\begin{tabular}{p{0.3\textwidth} p{0.17\textwidth} p{0.17\textwidth} p{0.17\textwidth}}

 & Simple & Approximate After-pulses & Front fit \\
\hline
Work time, $\%$ of acquisition time  & 61.1 & 158.8 & 9771.5 \\
\hline
Extracted, \% & 89.65 & 89.65 & 98.34 \\
\hline
Not extracted, \% & 8.96 & 8.9 & 3.48 \\
\hline
False positives,\%  & 0.01 & 0.01 & 2.37 \\
\hline
Effective dead time, $\mu s$ & 3.195 & 3.167 & 0.632 \\
\hline
\end{tabular}
\caption{Methods testing summary.}
\label{comparison-summary}
\end{table}

The first and second methods give the same quantitative results and have the same effective dead time of about 3 $\mu$s. The third method recognizes a lot more events and has 5 times less dead time compared to the first and second, but also has a much greater number of false positives and works 2 orders of magnitude slower.

\subsection{Discussion}
Of the three algorithms considered, "Front Fit" (number 3) shows the best dead time - 5 times less than "Simple" and "Approximate After-pulses". However, the nature of the frequent number of false positives remains in question. One of the reasons may be worse stability of the peak detection due to fitting on the front of the event. If the position will be extracted with error of 1-2 bin which is comparable to the size at which the after-pulse varies noticeably, the algorithm can increase next events amplitude error. In this case, it can be assumed that false events may have a different signal shape, which could be possibly classified and filtered out by neural network.

Another cause of false positives may be inadequate values of the fitted event parameters, which can also increase errors noise due to the after-pulse shifting. In addition, the current framework used for fitting is rather slow (it could be significantly optimized in future).

In the future, it is planned to analyze causes of false positives and improve the operation of the algorithm in both directions.

\subsection{Conclusions}
In this work, we considered ways to extract events from a continuous signal, taken from an ADC. We also considered an algorithm for creating a signal generator that simulates the operation of an ADC. The described extraction algorithms does not require additional knowledge about events and the spectrum and relies only on the acquired data. It makes possible to use them without significant changes for various signal digitization problems.

On the Lan10-12PCI board used, we could achieve a dead time of 0.6 $\mu$s, with a bin size of 320 ns.

The described algorithms were used to analyze data at the Troitsk nu-mass experiment on 2017\_05 and 2017\_11 data acquisition runs.

The implementation of the algorithms described in the article can be found in the repository on github \href{https://github.com/kapot65/signal-utils}{https://github.com/kapot65/signal-utils}.

The work was supported with RFBR grant “17-02-00361 A”. We would like to thank Vladislav Pantuev for manuscript revision and constructive criticism of the work.

\bibliographystyle{unsrt}
\bibliography{bibliography}

\end{document}